# Polycrystalline $\gamma$-boron: As hard as polycrystalline cubic boron nitride

Jiaqian Qin,[1*] Norimasa Nishiyama[1], Hiroaki Ohfuji[1], Toru Shinmei[1], Li Lei[1,2], Duanwei He[2], Tetsuo Irifune[1]

[1]*Geodynamics Research Center, Ehime University, Matsuyama, 790-8577, Japan*

[2]*Institute of Atomic and Molecular Physics, Sichuan University, Chengdu 610065, P.R. China*

The Vickers hardness of polycrystalline $\gamma$-B was measured using a diamond indentation method. The elastic properties of polycrystalline $\gamma$-B were determined using ultrasonic measurement method. Under the loading force up to 20 N, our test gave an average Vickers hardness in the asymptotic-hardness region of 30.3 GPa. We also measured the hardness and elastic properties of polycrystalline $\beta$-B and PcBN for comparison. The hardness and elastic properties for polycrystalline $\gamma$-B was found to be very close to that of PcBN.

[*] Corresponding author: E-mail: jiaqianqin@gmail.com

Boron is a chemically complex element. Among elemental boron polymorphs, only α-rhombohedral (α-B), β-rhombohedral (β-B), γ-orthorhombic (γ-B), β-tetragonal (T-192), and α-tetragonal boron (T-$B_{50}$ or T-$B_{52}$) have been currently established as pure phases [1-12]. The elemental neighbor to carbon in the periodic system of elements, boron is known to have polymorphs that are superhard [13-16] and all of its crystal structures are distinct from any other element. In particular, the γ-B has a hardness of above 50 GPa [13, 14]. This value makes the γ-B to be the second hardest elemental solid after diamond.

To date, a number of theoretical studies have been performed to investigate the mechanical properties of γ-B [17-19]. The calculated elastic moduli indicate that γ-B could be a superhard material with a Vicker's hardness as high as $B_6O$. However, the ideal shear strength result indicate that the failure mode in γ-B is dominated by the shear deformation type in the slip system (001)[010], and the weakest slip system has a relatively low ideal strength of about 21.6 GPa. In addition, the theoretical hardness of γ-B is about 42 GPa and 49 GPa according to Lyakhov's [20] and Chen's [21] hardness models. The measured hardness is usually lower than the ideal hardness (except for nano-polycrystalline), so these theoretical results are not consistent with the measured hardness of above 50 GPa [18]. Experimental and theoretical bulk moduli were also reported [14, 17, 22-25]. However, they didn't show the same bulk modulus ($B_0$=237 GPa [22,24] and 227 GPa [14,23]) in the *in situ* diamond anvil cell (DAC) experiment, because boron is a very weak scatterer of x-rays. And the first-principles also gave different value of the bulk modulus ($B_0$=222 GPa [14,17], 230.6 GPa [25] and 241 GPa [24]). Due to neglect the lattice vibrations, the previous results [14,17,25] are lower than Isaev *et al.* [24], calculated results. To verify the mechanical properties of polycrystalline γ-B, here we present the synthesis of nearly full densification polycrystalline $γ$-B compacts at high pressure and high temperature (HPHT). We also prepared polycrystalline $β$-B and PcBN (cBN-Ti$_3$SiC$_2$) compacts at HPHT for comparison. High-pressure experiments were carried out in kawai-type multi anvil apparatus and DS6×8MN cubic press, and the experimental details were described elsewhere [10, 26]. We investigated the elastic modulus of the polycrystalline $γ$-, $β$-B, and cBN-Ti$_3$SiC$_2$ compacts using an ultrasonic method. And the hardness was also tested with a larger applied load up to 20 N.

In this work, we prepared the nearly full densification polycrystalline $γ$-B compacts from $β$-B (claimed purity 99.6%, Goodfellow, and 99.9999%, ApexChem Boron) at 15 GPa and 1800 ºC for 30 min. We also sintered the nearly full densification polycrystalline $β$-B compacts at 6 GPa and 1700 ºC for 30 min. The recovered synthetic samples are all well sintered cylinder-shaped chunks that are about 2.5 mm in diameter. We have prepared two types of polycrystalline $γ$-B (OT1023 from Goodfellow Boron, and OT1025 from ApexChem Boron) and two types of polycrystalline $β$-B (OS2261 from Goodfellow boron, and OS2255 from ApexChem Boron). Both ends of these samples were polished with 1 $μ$m diamond paste and the sample was measured with a precision of 1 $μ$m. Sample lengths of the OT1023, OT1025, OS2261, and OS2255 are 1.861 mm, 1.891 mm, 2.247 mm, and 1.700 mm, respectively. From the scanning electron microscopy (SEM) observations it could be concluded that the samples exhibit a dense microstructure. Figure 1(a) shows the SEM image of polycrystalline $γ$-B, SEM image indicate that the grain size is about 1-30 $μ$m, and the sample has very tight microstructure. X-ray diffraction confirmed that the samples are pure $γ$- and $β$-B [Figure 1(b)], respectively. The densities of the sintered samples were measured using the Archimedes method, which gives a value of 2.48±0.02 g/cm$^3$ and 2.29±0.02 g/cm$^3$ for polycrystalline $γ$- and $β$-B,

98% and 97% of the theoretical density [27, 28], respectively. The polycrystalline $\gamma$- and $\beta$-B were then examined by ultrasonic measurement and hardness test.

Elastic property measurements were carried out using ultrasonic technique at ambient condition. The data on the measurement of the velocity of ultrasonic waves transition through the samples were collected via ultrasonic measurement using a pulse reflection method. Both longitudinal and transverse wave signals were generated and received by a 10°Y-cut LiNbO$_3$ transducer. The LiNbO$_3$ transducer was attached at the end of a WC block, and the sample was mounted on the opposite of the WC block. The experimental details were described elsewhere [29]. The longitudinal and transverse wave travel times ($t$) through the sample determined by comparing the arrival times of two signals reflected at the front ($R_0$) and back ($R_1$) ends of the sample. Thus, the wave velocity ($v$) can be calculated by dividing the round-trip distance by $t$. Values for the Young's modulus ($E$), bulk modulus ($B$), shear modulus ($G$), and Poisson's ratio ($v$) could be calculated from the densities of the sintered bodies together with longitudinal ($V_L$) and transverse ($V_T$) wave velocities [30].

Table 1 lists the raw materials, longitudinal ($V_L$) and transverse ($V_T$) of the polycrystalline boron and the calculated bulk modulus ($B$). We obtained the very same $V_L$, $V_T$ and $B$ using different starting materials for polycrystalline $\gamma$- and $\beta$-B, respectively. These results give us confidence in the reliability of ultrasonic measurement for calculating the elastic modulus of polycrystalline $\gamma$- and $\beta$-B. Table 2 presents our results of Young's modulus ($E$), bulk modulus ($B$), shear modulus ($G$), and Poisson's ratio ($v$) of polycrystalline $\gamma$- and $\beta$-B, respectively, in comparison with previous experimental and calculated results [2,5,14,17,19,22,24,25,31-33]. In table 2, the results show that the elastic modulus of $\gamma$-B is always higher than that of $\beta$-B, which is in agreement with the previous results. On the other hand, we find that our ultrasonic measurement results are lower than the calculated and DAC results. The porosity in the polycrystalline γ- and β-B might underestimate the elastic modulus. The measured density of our polycrystalline γ- and β-B are within 98% and 97% of the theoretical density, indicating approximate 2% and 3% of porosity in the polycrystalline γ- and β-B samples, respectively. Porosity at some level inevitably exists in real materials. Based on the previous results [34], we estimate the bulk modulus can be reduced by about 5-10% in the presence of 2% and 3% porosity in our polycrystalline specimen. Therefore, the bulk and shear modulus of the full densification are 225-237 GPa and 239-252 GPa, respectively. Isaev *et al.* [24] performed accurate measurements and *ab initio* calculations of the EOS for the γ-B up to 40 GPa including the effect of lattice vibrations. They obtained the experimental and calculated bulk modulus is higher than other results, but it is in good agreement with the present determination. And they also demonstrated that the phonon contribution had a profound impact on the EOS of γ-B, leading to a substantially reduced value of B′ that greatly improves the agreement between theory and the experimental values.

Compared with boron suboxide ($B_6O$) (B=222 GPa [35] and G=204 GPa [36]), γ-B has a very similar bulk modulus; moreover, its shear modulus is much higer than that of $B_6O$. According to Teter, [37] the polycrystalline shear modulus is a better predictor of hardness. Because $B_6O$ is a well-known superhard material with a Vicker's hardness as high as 45 GPa, [38] it is conceivable that γ-B is also a superhard material considering its higher shear modulus than that of $B_6O$. Additionally, the hardness of γ-B should higher than that of β-B. However, our previous experiences indicate that the high bulk and shear modulus materials are not always superhard materials. Thus, to further explore the hardness nature, the Vickers hardness of the polycrystalline boron was performed by means of a Vickers indentation method using a pyramidal diamond indenter. The loading force of the hardness tester could be adjusted from 1.96 to 19.6 N (1.96, 2.94, 4.9, 9.8, and 19.6 N loads). The dwelling time was fixed at 15 s. Under each load, four or five indentations were made. Under a certain applied load of *P*, *Hv* was determined by $Hv=1854.4P/d^2$, where *d* is the arithmetic mean of the two diagonals of the indent in micrometers [26]. The averaged *Hv* values were 35.51±0.09, 32.63±0.01, 31.14±0.01, 30.96±0.02, 30.31±0.01 GPa under a load of 1.96, 2.94, 4.9, 9.8, and 19.6 N, respectively. The results have showed that the hardness appears to increase with a decrease in load. In order to compare hardness values of different materials, we measured *Hv* of polycrystalline *β*-B compacts. The hardness test gave values of 30.54±0.04, 29.76±0.04, 27.09±0.01, 26.71±0.02 GPa under a load of 1.96, 2.94, 4.9, and 9.8 N, respectively. Figure 2 summarizes the average measured *Hv* for polycrystalline *γ*- and *β*-B under different loads, and the inset shows the Vickers indentations of polycrystalline *γ*-B produced by a diamond pyramidal indenter under different load. The hardness results indicate that the tendency of hardness to decrease becomes weak for large loads, and the hardness of polycrystalline *γ*-B is always higher than that of polycrystalline *β*-B under every load, which is in good agreement with the elastic modulus results. Since the polycrystalline *γ*-B is not fully densification sample, if we consider the effect of porosity, the hardness of fully densification polycrystalline *γ*-B should be above 30.54 GPa. This result will be very close to the theoretical hardness (42 GPa) according to Lyakhov's [20] hardness models. However, the present hardness results are not consistent with the previous reported Vickers hardness of 50-58 GPa [13, 14], but it is in good agreement with the theoretical work [18]. Zhou, *et al*. [18] investigated the stress-strain relations of γ-B under shear deformation, and found that the failure mode is dominated by the shear deformation type in the slip system (001)[010]. The weakest slip system (001)[010] has relatively low ideal shear strength of about 21.6 GPa, which is not consistent with the measured hardness of above 50 GPa. So we can find that our hardness is very consistent with Lyakhov's [20]. hardness models and the ideal shear strength result [18]. According to the grain size of present sample is micron scale and we measured the hardness under a larger loading force, we can explain the difference between our hardness results and the previous reported. Firstly, the experimental Vickers micro hardness test is complicated, since they were carried out submicron-sized grain or polycrystalline samples where contributions from different crystal orientations are expected. Consequently, Solozhenko, *et al.* [13] obtained the large dispersion of the measured hardness values in Figure 1(b) of Ref. [13], and a Vickers hardness of 40-55 GPa under 5 N loading force, which is consistent with the tensile and shear strength of 21.6 GPa-65 GPa [18]. Secondly, Zarechnaya, *et al.* [14] reported the hardness of 105 GPa to 58 GPa with 0.5 N-10 N loading forces, which is not consistent with the theoretical work [18]. The reported very hardness might duo to the nanosized nature, especially for the crystallite size of 10-15 nm is within the range of the "strongest size", such as Nano polycrystalline diamond [39] and Nano polycrystalline cubic boron nitride [40]. Furthermore, the radial crack length of Vickers indentation has been directly measured using optical microscopy. Therefore, the fracture toughness $K_{IC}$ can be calculated using the experimentally determined values of crack length and Young's modulus. The fracture toughness $K_{IC}$ is 4.1 and 3.3 MPa m$^{1/2}$ for polycrystalline γ- and β-B, respectively. The fracture toughness of polycrystalline γ-B is very close to that of diamond (5 MPa m$^{1/2}$) [41], and cBN (2.8 MPa m$^{1/2}$) [42].

In order to compare hardness values of different materials and estimate the credibility of our hardness and elastic properties data, we also prepared the cBN-10wt.% $Ti_3SiC_2$ composites (PcBN) at 5 GPa and 1400 ºC for 30 min, and measured the hardness and elastic properties of PcBN. Table 3 presents the hardness, Young's modulus ($E$), bulk modulus ($B$), shear modulus ($G$), and B/G value of polycrystalline $γ$- and $β$-B and PcBN, compared with previous PcBN, PCD, and $B_6O$-$B_4C$ composites [29, 43, 44]. The measured hardness and elastic properties of PcBN are in agreement with values given in the Ref. [43], which further demonstrates the reliability of our experimental results. In the Table 3, we can find the hardness and elastic properties of polycrystalline $γ$-B are very close to that of PcBN and $B_6O$-$B_4C$ composites. Additionally, Pugh proposed the ratio of bulk to shear modulus ($B/G$) as an indication of ductile *vs* brittle characters [45]. The bulk modulus $B$ represents the resistance to fracture, while the shear modulus $G$ represents the resistance to plastic deformation. A low $B/G$ ratio is associated with brittle, whereas a high value corresponds to ductility nature. If $B/G$>1.75, the material behaves in a ductile manner; otherwise, the material behaves in a brittle manner. In the case of polycrystalline $γ$-B, our experimental $B/G$ value is 0.941, which is smaller than the $B_6O$-$B_4C$ composites and $β$-B, but it is very similar with that of PCD and PcBN values. Thus, it is conceivable that polycrystalline $γ$-B is also a superhard polycrystalline material.

In summary, we have measured the elastic properties and hardness of polycrystalline boron. Our hardness and previous theoretical results indicate that the hardness of $γ$-$B_{28}$ cannot reach more than 50 GPa, and suggest that the reported very hardness is duo to the different crystal orientations and nanosized nature. We find that polycrystalline $γ$-B possesses a similar elastic moduli and hardness with PcBN and

B$_6$O compacts. These results suggest that polycrystalline $\gamma$-B$_{28}$ can also be a superhard polycrystalline material, and expect that it can be widely used.

This work is supported by JSPS Grants-in Aid for Scientific Research under Grant No. 22 00029 and G-COE Program. One of authors (J. Qin) would like to thank Regshi Ryo and Yongtao Zou for helping in experiment.

**Figure Caption**

Figure 1. (a) SEM images of polycrystalline γ-B, (b) X-ray diffraction of polycrystalline γ- and β-B.

Figure 2. Summarize the average measured *Hv* for polycrystalline γ- and β-B under different loads, and the inset shows the Vickers indentations of polycrystalline γ-B produced by a diamond pyramidal indenter under different load. The hardness results indicate that the tendency of hardness to decrease becomes weak for large loads, and the hardness of polycrystalline γ-B is always higher than that of polycrystalline β-B under each load.

**Table Caption**

Table 1. List of the starting materials, longitudinal ($V_L$) and transverse ($V_T$) of the polycrystalline boron and the calculated bulk modulus (*B*)

Table 2. Present our results of Young's modulus (*E*), bulk modulus (*B*), shear modulus (*G*), and Poisson's ratio (*υ*) of polycrystalline γ- and β-B, respectively, in comparison with previous experimental and calculated results [2, 6, 14, 17, 19, 22, 24, 25, 31-33].

Table 3. Show the hardness, Young's modulus (*E*), bulk modulus (*B*), shear modulus (*G*), and B/G value of polycrystalline γ- and β-B and cBN-10wt.% $Ti_3SiC_2$ composites, compared with previous PcBN, PCD, and $B_6O$-$B_4C$ composites [29,43,44].